# Computational forms for binary particle interactions at different levels of anisotropy

Christopher N. Everett ★ and Garret Cotter

*Astrophysics Sub-department, Department of Physics, University of Oxford, Denys Wilkinson Building, Keble Road, Oxford OX1 3RH, UK*

**ABSTRACT**
Particle interactions are key elements of many dynamical systems. In the context of systems described by a Boltzmann equation, such interactions may be described by a collision integral, a multidimensional integral over the momentum-phase space of the interaction. This integral is often simplified by assuming isotropic particle distributions; however, such an assumption places constraints on the dynamics of the system. This paper presents computational forms of the collision integral for relativistic, binary interactions at three levels of anisotropy, including a novel form in the isotropic case. All these forms are split into two parts, an absorption and an emission spectrum, which may be pre-calculated via numerical integration for simulation purposes. We demonstrate the use of these forms by comparison with the analytically integrated, isotropic emission spectrum of electron–positron annihilation, which are shown to agree to numerical precision. The emission spectrum is then further extended to axisymmetric particle distributions, where two-dimensional spectral maps can be generated to provide new insight.

## 1 INTRODUCTION

In the context of high-energy astrophysical sources, accurate modelling of processes, such as electron–positron pair annihilation/creation, Compton scattering, synchrotron, etc., is critical. Prior works on the emission and absorption spectra of such processes have focused mainly on isotropic interactions (Weaver 1976; Dermer 1984, 1985, 1986; Baring 1987; Coppi & Blandford 1990), with only small ventures into the world of anisotropy (Stepney & Guilbert 1983). The assumption of isotropy in these works is often a compromise between simplicity and complexity as the dimensionality of the integrals rapidly increases with anisotropy.

An example application is the emission modelling of astrophysical jets, where the current state-of-the-art models (Potter 2018; Cerruti et al. 2021; Lucchini et al. 2022; Zacharias et al. 2022; Klinger et al. 2023) evolve shells of isotropic non-thermal plasma travelling at a bulk Lorentz factor out from the central black hole. The constraint of isotropy prevents self-consistent 'hydrodynamical' evolution of the jets as plasma shells cannot interact with each other. In addition, isotropy is only a good approximation if the dynamical timescales of the system are much greater than isotropizing time-scales, which is not guaranteed at all energies. Work in the field of pair-plasma winds (Aksenov, Milgrom & Usov 2004; Aksenov, Ruffini & Vereshchagin 2007, 2010) has demonstrated self-consistent evolution of anisotropic particle distributions using a kinetic approach[1] showing that the additional dimensionality associated with anisotropy is within the grasp of modern computing. However, a clear framework in which to treat general anisotropic interactions is missing in the literature.

Although inspired by astrophysical sources, the work presented here is applicable to a wide range of physical systems where reactions, collisions, and interactions play an important role. To that aim, effort has been made to keep the forms of the emission and absorption spectra given in this paper as general as possible, while presenting them in a way that is easy for the reader to apply to any desired interaction. To maintain a reasonable scope for this paper, the forms developed here pertain to a specific subset of interactions, though may be easily extended upon in future works. The subset consists of all binary interactions that are reversible, i.e. take the form $12 \rightleftharpoons 34$. No assumptions are made about the species of particles involved in the interactions, i.e. they can be massive or massless, nor about their energy/momenta; therefore, the forms given in this papers are relativistic and may be evaluated in an arbitrary frame of reference.

The format of the rest of this paper is as follows. In Section 2, a general overview, context, and definitions of reaction rates, emission, and absorption spectra are provided. In Section 3, a framework in which to treat systems that exhibit axisymmetry in their particle distributions is developed, and then Section 4 examines the isotropic cases. Section 5 identifies and examines some singular cases of the integrals developed in the previous two sections that are of particular note and develops new forms dedicated to these cases. In Section 6, an example application of this work to electron–positron annihilation is given, by comparison between the computational forms developed in this work and the known analytical form of the isotropic emissivity. The comparison demonstrates the replicability of the isotropic emissivity without need for difficult and time-consuming algebraic integration, and hence may be applied to a wider range of problems. The comparison is then further expanded by examining the axisymmetric emission spectrum for two test cases, an axisymmetric case and a cosmic ray case – generating a two-dimensional (2D) spectral map of the emissivity for each that provides novel insight.

---

★ E-mail: christopher.everett@physics.ox.ac.uk

[1] Treatment of integration bounds and domain conditions of the emission and absorption terms tends to be rather ambiguous; however, their approach appears to be consistent with this work's equations (12) and (13) if integrated over $\boldsymbol{p}_1$ momenta.

## 2 OVERVIEW

The dynamics of a system that involves interaction between particles may generally (neglecting external forcing) be described in kinetic theory by a set of relativistic Boltzmann equations (one for each species in the system):

$$p^\mu \partial_\mu f(x^\mu, \boldsymbol{p}) = C(x^\mu, \boldsymbol{p}), \tag{1}$$

where $x^\mu = (t, \boldsymbol{x})$ is the four-displacement, $p^\mu = (p^0, \boldsymbol{p})$ is the four-momentum,[2] $C(x^\mu, \boldsymbol{p})$ is the collision integral, and $f(x^\mu, \boldsymbol{p})$ is the distribution function defined by $\int f(x^\mu, \boldsymbol{p}) \mathrm{d}^3 \boldsymbol{x} \mathrm{d}^3 \boldsymbol{p} = N$, the number of particles. The on-shell condition relating energy to momentum $m^2 = (p^0)^2 - \boldsymbol{p} \cdot \boldsymbol{p}$ constrains $f$ such that it is only a function of time, position, and momentum, not energy. Henceforth, the space–time positional dependence $x^\mu$ of terms will be dropped for brevity.

The collision integral $C(\boldsymbol{p}_1)$[3] is itself an integral over momentum states (hence its name) involving the product of distribution functions $f(\boldsymbol{p}_i)$ and transition rates $W$ (Groot, Leeuwen & Weert 1980; Cercignani 2002):[4]

$$C(\boldsymbol{p}_1) = \int \frac{\mathrm{d}^3 \boldsymbol{p}_2}{p_2^0} \frac{\mathrm{d}^3 \boldsymbol{p}_3}{p_3^0} \frac{\mathrm{d}^3 \boldsymbol{p}_4}{p_4^0} \left[ \frac{f(\boldsymbol{p}_3) f(\boldsymbol{p}_4)}{1 + \delta_{34}} W(p_3^\mu, p_4^\mu | p_1^\mu, p_2^\mu) \right.$$
$$\left. - \frac{f(\boldsymbol{p}_1) f(\boldsymbol{p}_2)}{1 + \delta_{34}} W(p_1^\mu, p_2^\mu | p_3^\mu, p_4^\mu) \right], \tag{2}$$

$$W(p_1^\mu, p_2^\mu | p_3^\mu, p_4^\mu) = \frac{s p_{\text{in}}^{*2}}{\pi} \sigma_{12|34}(s, t) \delta^{(4)}(p_1^\mu + p_2^\mu - p_3^\mu - p_4^\mu), \tag{3}$$

where $\delta^{(4)}$ is the 4D Dirac delta function, $\sigma_{12|34}(s, t) \equiv \frac{\mathrm{d}\sigma_{12|34}}{\mathrm{d}t}$ is the Lorentz-invariant differential cross-section, which for a binary interaction is only dependent on the Mandelstam variables $s, t$ defined in Appendix A, and $p_{\text{in}}^*$ is the magnitude of the incoming particle momentum in the centre of momentum frame. In the case of the forward interaction $12 \to 34$,

$$p_{\text{in}}^* = p_1^* = p_2^* = \frac{\sqrt{\lambda(s, m_1^2, m_2^2)}}{2\sqrt{s}} = \frac{\mathcal{F}_{12}(s)}{\sqrt{s}}, \tag{4}$$

where $\lambda(s, m_1^2, m_2^2)$ is Källén's function (Källén 1964) and $\mathcal{F}_{12}(s)$ is the invariant flux of incoming particles. The additional factor of $\frac{1}{1+\delta_{34}}$ has been included in equation (2) so as not to overcount $(\boldsymbol{p}_3, \boldsymbol{p}_4)$ momentum states when the particles are indistinguishable (see Cercignani 2002, section 2.1 for a discussion). A similar factor accounting for the indistinguishability of $(\boldsymbol{p}_1, \boldsymbol{p}_2)$ states is not needed as this cancels with the 'true' collision integral for that particle being $C(\boldsymbol{p}_1) + C(\boldsymbol{p}_2)$. Integrating $C(\boldsymbol{p}_1)$ over the Lorentz invariant momentum-space volume element $\frac{\mathrm{d}^3 \boldsymbol{p}_1}{p_1^0}$ is equal to the reaction rate $R_1$, which is the *net* rate of change of particles of type 1 per volume of space:

$$R_1 = \int \frac{\mathrm{d}^3 \boldsymbol{p}_1}{p_1^0} C(\boldsymbol{p}_1). \tag{5}$$

Performing the same integral over equation (1) yields

$$(\partial_t + \boldsymbol{v} \cdot \boldsymbol{\nabla}) n = R, \tag{6}$$

where $n = \int f(\boldsymbol{p}) \mathrm{d}^3 \boldsymbol{p}$ is the number density and $\boldsymbol{v}$ is the bulk velocity. The aforementioned form may be more familiar than equation (1) particularly when the nature of the distribution of particles over momentum space is not needed or relevant to the system.

For practical purposes and to align with previous works, the collision spectrum is defined here as

$$C_{\text{spe}}(\boldsymbol{p}) \equiv \frac{C(\boldsymbol{p})}{p^0}. \tag{7}$$

Following the notation of Baring (1987), the collision spectrum splits into two portions:

$$C_{\text{spe}}(\boldsymbol{p}) = S_{\text{spe}}(\boldsymbol{p}) - T_{\text{spe}}(\boldsymbol{p}). \tag{8}$$

The absorption spectrum

$$T_{\text{spe}}(\boldsymbol{p}_1) = \frac{1}{1 + \delta_{34}} \frac{1}{p_1^0} \int \frac{\mathrm{d}^3 \boldsymbol{p}_2}{p_2^0} \frac{\mathrm{d}^3 \boldsymbol{p}_3}{p_3^0} \frac{\mathrm{d}^3 \boldsymbol{p}_4}{p_4^0}$$
$$\times f(\boldsymbol{p}_1) f(\boldsymbol{p}_2) W(p_1^\mu, p_2^\mu | p_3^\mu, p_4^\mu), \tag{9}$$

accounts for the *loss* of type-1 particles with particular momentum $\boldsymbol{p}_1$ due to the forward reaction $12 \to 34$ (also referred to as the 'loss term' in literature), while the emission spectrum

$$S_{\text{spe}}(\boldsymbol{p}_1) = \frac{1}{1 + \delta_{34}} \frac{1}{p_1^0} \int \frac{\mathrm{d}^3 \boldsymbol{p}_2}{p_2^0} \frac{\mathrm{d}^3 \boldsymbol{p}_3}{p_3^0} \frac{\mathrm{d}^3 \boldsymbol{p}_4}{p_4^0}$$
$$\times f(\boldsymbol{p}_3) f(\boldsymbol{p}_4) W(p_3^\mu, p_4^\mu | p_1^\mu, p_2^\mu), \tag{10}$$

accounts for the *gain* of type-1 particles into a particular state $\boldsymbol{p}_1$ from the reverse reaction $34 \to 12$ (also referred to as the 'gain term').

The absorption integral does not depend on $\boldsymbol{p}_3, \boldsymbol{p}_4$ apart from in the transition rate equation (3), which can integrated over to give

$$\int \frac{W(p_1^\mu, p_2^\mu | p_3^\mu, p_4^\mu)}{(1 + \delta_{34}) p_3^0 p_4^0} \mathrm{d}^3 \boldsymbol{p}_3 \mathrm{d}^3 \boldsymbol{p}_4 = \mathcal{F}_{12}(s) \sigma_{12|34}(s), \tag{11}$$

where $\sigma_{12|34}(s)$ is the cross-section for the interaction $12 \to 34$.[5] Substituting this in equation (10) gives yields the simple form

$$T_{\text{spe}}(\boldsymbol{p}_1) = \frac{1}{p_1^0} \int \frac{\mathrm{d}^3 \boldsymbol{p}_2}{p_2^0} f(\boldsymbol{p}_1) f(\boldsymbol{p}_2) \mathcal{F}_{12}(s) \sigma_{12|34}(s). \tag{12}$$

As $s = s(p_1^\mu, p_2^\mu)$ this is a form that, in theory, can be integrated depending on the nature of the distribution functions.

The emission integral is less simple as now there is dependence on $\boldsymbol{p}_3, \boldsymbol{p}_4$ via their distribution functions and hence the simplification of equation (11) cannot be used. Substituting equation (3) into equation (9), noting $p_{\text{in}}^* = \frac{\mathcal{F}_{34}(s)}{\sqrt{s}}$ as this is the reverse reaction, and

---

[2] The quantity $p^0 = E/c$, i.e. has dimensions of momentum, with $c = 1$ and a metric signature $(+ - - -)$ being used throughout this text.
[3] Numbered subscripts refer to particles in the binary interaction $12 \rightleftharpoons 34$, which may be of the same species.
[4] For the interactions relevant to this work, all distributions are assumed to have Boltzmann statistics.

[5] Readers should take care as the factor of $1/(1 + \delta_{34})$ is typically, but not always, absorbed into definitions of the total cross-section found in the literature depending on whether the collision can be treated classically or quantum mechanically.

integrating over $\boldsymbol{p}_2$

$$S_{\text{spe}}(\boldsymbol{p}_1) = \frac{1}{1+\delta_{34}} \frac{2}{\pi p_1^0} \int \frac{d^3 \boldsymbol{p}_3}{p_3^0} \frac{d^3 \boldsymbol{p}_4}{p_4^0} f(\boldsymbol{p}_3) f(\boldsymbol{p}_4) \mathcal{F}_{34}^2(s)$$
$$\times \sigma_{34|12}(s,t) \delta\left(s+t+u-\sum_i m_i^2\right) \Theta\left(p_3^0 + p_4^0 - p_1^0\right), \quad (13)$$

where $\Theta$ is the Heaviside step function.

Equations (12) and (13) are the simplest general forms of the emission and absorption spectrum when the distribution functions are assumed to be fully anisotropic in momentum space.[6]

## 3 AXISYMMETRIC MOMENTUM SPACE

Systems which exhibit a *global* preferred direction, e.g. $\hat{\boldsymbol{r}}$ for a spherically expanding shock or $\boldsymbol{B}$ in a magnetized flow, permit their distribution functions to be described as *locally*[7] axisymmetric in momentum space. Local axisymmetry occurs when any momentum-orthogonal forcing term, e.g. $\boldsymbol{v} \times \boldsymbol{B}$, averages to zero on relevant time- and/or length-scales.

Working in spherical coordinates for momentum space $(p, \theta, \phi)$, where $p = |\boldsymbol{p}|$, momentum-space axisymmetry about the polar axis yields that all properties derived from the distribution function are independent of $\phi$. With this property, it is convenient to replace the distribution functions $f(\boldsymbol{p})$, absorption spectrum $T_{\text{spe}}(\boldsymbol{p})$, and emission spectrum $S_{\text{spe}}(\boldsymbol{p})$, with their axisymmetric counterpart e.g. $f(p, \theta) = 2\pi p^2 f(\boldsymbol{p})$, which equates to integrating over the momentum-space element $p^2 d\phi$. In other words, $\int f(\boldsymbol{p}) d^3 \boldsymbol{p} = \int f(p, \theta) dp d\cos\theta$, hence $f(p, \theta)$ equal to a constant describes a distribution of particles whose density per unit energy is constant, whereas $f(\boldsymbol{p})$ being a constant describes a distribution that increases in density with energy.

### 3.1 Axisymmetric absorption spectrum

Under the assumption of axisymmetry, the absorption spectrum (equation 12) has no dependence on $\phi_1, \phi_2$ apart from the terms containing $s$. A change of variables can be instated from $d\phi_2 \to ds$, to decouple $d\phi_1$, then integrating equation (12) over $p_1^2 d\phi_1$ yields the axisymmetric absorption spectrum $T_{\text{spe}}(p_1, \theta_1)$:

$$T_{\text{spe}}(p_1, \theta_1) = \frac{f(p_1, \theta_1)}{2\pi p_1^0} \int \frac{dp_2 d\cos\theta_2}{p_2^0} f(p_2, \theta_2)$$
$$\times \int_{s_-}^{s_+} \frac{ds}{\sqrt{(s_+ - s)(s - s_-)}} \mathcal{F}_{12}(s) \sigma_{12|34}(s), \quad (14a)$$

with

$$s_\pm = m_1^2 + m_2^2 + 2p_1^0 p_2^0 - 2p_1 p_2 \cos(\theta_2 \pm \theta_1). \quad (14b)$$

The $s$ integration bounds in equation (14) must be further constrained by the physical region of the Mandelstam variables (see Appendix A), in this case there is only one relevant condition which is equation (A5).

When the particle species are restricted to be massless, this form is equivalent to Stepney & Guilbert (1983, equation B5), when integrated over $dp_1 d\cos\theta_1$.

### 3.2 Axisymmetric emission spectrum

Under the assumption of axisymmetry, the emission spectrum (equation 13) has no dependence on $\phi_1, \phi_3, \phi_4$ apart from the terms containing $s$, $t$ and $u$. Integrating over $p_1^2 d\phi_1$ and changing variables from $d\phi_1 d\phi_3 d\phi_4 \to ds dt du$ gives

$$S_{\text{spe}}(p_1, \theta_1) = \frac{1}{1+\delta_{34}} \frac{p_1^2}{2\pi^3 p_1^0} \int \frac{dp_3 dp_4}{p_3^0 p_4^0} \int d\cos\theta_3 d\cos\theta_4$$
$$\times f(p_3, \theta_3) f(p_4, \theta_4) \int_{s_-}^{s_+} ds \frac{\mathcal{F}_{34}^2(s) \Theta\left(p_3^0 + p_4^0 - p_1^0\right)}{\sqrt{(s_+ - s)(s - s_-)}}$$
$$\times \int_{t_-}^{t_+} \int_{u_-}^{u_+} dt du \frac{\sigma_{34|12}(s,t) \delta\left(s+t+u-\sum_i m_i^2\right)}{\sqrt{(t_+ - t)(t - t_-)} \sqrt{(u_+ - u)(u - u_-)}}. \quad (15)$$

The delta function can be integrated over using the $s, t,$ or $u$ integrals. Choosing the $u$ integral

$$S_{\text{spe}}(p_1, \theta_1) = \frac{1}{1+\delta_{34}} \frac{p_1^2}{2\pi^3 p_1^0} \int \frac{dp_3 dp_4}{p_3^0 p_4^0} \int d\cos\theta_3 d\cos\theta_4$$
$$\times f(p_3, \theta_3) f(p_4, \theta_4) \int_{s_-}^{s_+} ds \frac{\mathcal{F}_{34}^2(s)}{\sqrt{(s_+ - s)(s - s_-)}}$$
$$\times \int_{t_-}^{t_+} dt \frac{\sigma_{34|12}(s,t)}{\sqrt{(t_+ - t)(t - t_-)}} \frac{\Theta\left(p_3^0 + p_4^0 - p_1^0\right)}{\sqrt{(u_+ - u)(u - u_-)}}, \quad (16a)$$

with

$$s_\pm = m_3^2 + m_4^2 + 2p_3^0 p_4^0 - 2p_3 p_4 \cos(\theta_3 \pm \theta_4), \quad (16b)$$

$$t_\pm = m_3^2 + m_1^2 - 2p_3^0 p_1^0 + 2p_3 p_1 \cos(\theta_3 \mp \theta_1), \quad (16c)$$

$$u_\pm = m_4^2 + m_1^2 - 2p_4^0 p_1^0 + 2p_4 p_1 \cos(\theta_4 \mp \theta_1), \quad (16d)$$

and the $u$ is given by equation (A4). When integrating the delta function, the domain of $u$ integral dictates that the emission spectrum is non-zero only for

$$u_- \leq u \leq u_+. \quad (16e)$$

Additional constraints on the Mandelstam variables, described in Appendix A, also apply, in particular equations (A5–A8).

## 4 ISOTROPIC MOMENTUM SPACE

Systems may be described as *locally* isotropic in momentum space if all forcing terms may be averaged to zero and sufficient collisions have occurred on relevant time- and/or length-scales.

Again with working spherical momentum-space coordinates, isotropy is described by distribution functions that are independent of $\theta$ and $\phi$. With this property it is convenient to replace the distribution functions $f(\boldsymbol{p})$, absorption spectrum $T_{\text{spe}}(\boldsymbol{p})$, and emission spectrum $S_{\text{spe}}(\boldsymbol{p})$ with their isotropic counterparts, e.g. $f(p) = 4\pi p^2 f(\boldsymbol{p})$, which equates to integrating over the momentum-space element $p^2 \sin\theta d\theta d\phi = p^2 d\Omega$.

---

[6] In principle, the delta function can be used to integrate over one of the six integrals involved. This approach has not been chosen, as the choice of integral to eliminate is situation dependent. However, Appendix B provides a useful form, whereby the emission spectrum is integrated over $p_1$ in order to remove the delta function.

[7] Locally is taken to refer to a local spatial axis; e.g. the distribution of particles from a spherically symmetric implosion may be described as being locally axisymmetric about the local radial direction but requires a connection coefficient to translate the momentum-space coordinates from one spatial point to another.

### 4.1 Isotropic absorption spectrum

Under the assumption of isotropy, equation (12) then has no dependence on $\theta_1, \theta_2, \phi_1, \phi_2$ apart from the terms containing $s$. Integrating over $p_1^2 d\Omega_1$ yields the isotropic absorption spectrum $T_{\text{spe}}(p_1)$:

$$T_{\text{spe}}(p_1) = \int d\Omega_1 \frac{f(p_1)}{4\pi p_1^0} \int \frac{d^3 \boldsymbol{p}_2}{p_2^0} \frac{f(p_2)}{4\pi p_2^0} \mathcal{F}_{12}(s) \sigma_{12|34}(s). \quad (17)$$

The angular variables may be freely rotated to a primed frame aligned to $\boldsymbol{p}_1$ with $d\Omega_1' d\Omega_2' = d\Omega_1 d\Omega_2$. Furthermore, in this frame $s = m_1^2 + m_2^2 + 2p_1^0 p_2^0 - 2p_1 p_2 \cos \theta_2'$. Making the change of variables $d\theta_2' \to ds$ decouples $d\theta_1', d\phi_1', d\phi_2'$. Integrating over these decoupled components gives the simplest form:

$$T_{\text{spe}}(p_1) = \frac{f(p_1)}{4 p_1 p_1^0} \int dp_2 \frac{f(p_2)}{p_2 p_2^0} \int_{s_-}^{s_+} ds \mathcal{F}_{12}(s) \sigma_{12|34}(s), \quad (18a)$$

with

$$s_\pm = m_1^2 + m_2^2 + 2 p_1^0 p_2^0 \pm 2 p_1 p_2, \quad (18b)$$

and the condition equation (A5) still applying.

This result can be made identical to Baring (1987, equation 27) with the further substitutions $\varepsilon^{*2} = s/4$ and $f(p)dp = nF(p^0)dp^0$.

### 4.2 Isotropic emission spectrum

Under isotropy, the emission spectrum equation (13) has no dependence on $\theta_3, \theta_4, \theta_1, \phi_3, \phi_4, \phi_1$ apart from the terms containing $s, t$ and $u$. Integrating over momentum-space solid angle $p_1^2 d\Omega_1$ yields the isotropic emission spectrum:

$$S_{\text{spe}}(p_1) = \frac{1}{1+\delta_{34}} \frac{p_1^2}{8\pi^3 p_1^0} \int \frac{dp_3 dp_4}{p_3^0 p_4^0} f(p_3) f(p_4)$$

$$\times \int d\Omega_1 d\Omega_3 d\Omega_4 \mathcal{F}_{34}^2(s) \sigma_{34|12}(s, t)$$

$$\times \delta(s + t + u - m_1^2 - m_2^2 - m_3^2 - m_4^2) \Theta\left(p_3^0 + p_4^0 - p_1^0\right). \quad (19)$$

Akin to the absorption spectrum (Section 4.1), the angular variables may be rotated to a primed frame where they are aligned to $\boldsymbol{p}_3$ with $d\Omega_1 d\Omega_3 d\Omega_4 = d\Omega_1' d\Omega_3' d\Omega_4'$. Furthermore, in this frame

$$s = m_3^2 + m_4^2 + 2 p_3^0 p_4^0 - 2 p_3 p_4 \cos \theta_4', \quad (20)$$

$$t = m_1^2 + m_3^2 - 2 p_1^0 p_3^0 + 2 p_1 p_3 \cos \theta_1', \quad (21)$$

$$u = m_1^2 + m_4^2 - 2 p_1^0 p_4^0 + 2 p_1 p_4 (\cos \theta_1' \cos \theta_4' + \cos \Phi \sin \theta_1' \sin \theta_4'), \quad (22)$$

where $\Phi = \phi_4' - \phi_1'$. Making the change of variables $d\theta_1' d\theta_4' d\phi_4' \to dt ds du$, decouples $\Omega_3'$ and $\phi_1'$ from the system and the delta function can be integrated over to remove $u$ dependence and yield:

$$S_{\text{spe}}(p_1) = \frac{1}{1+\delta_{34}} \frac{p_1}{4\pi p_1^0} \int \frac{dp_3 dp_4}{p_3^2 p_4 p_3^0 p_4^0} f(p_3) f(p_4)$$

$$\times \int_{s_-}^{s_+} ds \, \mathcal{F}_{34}^2(s) \int_{t_-}^{t_+} dt \, \frac{\sigma_{34|12}(s, t)}{\sqrt{(u_+ - u)(u - u_-)}} \Theta\left(p_3^0 + p_4^0 - p_1^0\right), \quad (23a)$$

where

$$s_\pm = m_3^2 + m_4^2 + 2 p_3^0 p_4^0 \pm 2 p_3 p_4, \quad (23b)$$

$$t_\pm = m_3^2 + m_1^2 - 2 p_3^0 p_1^0 \pm 2 p_3 p_1, \quad (23c)$$

$$u_\pm = m_1^2 + m_4^2 - 2 p_1^0 p_4^0 + 2 p_1 p_4 (\cos \theta_1' \cos \theta_4' \pm \sin \theta_1' \sin \theta_4'), \quad (23d)$$

with

$$\cos \theta_1' = \frac{2t - t_+ - t_-}{4 p_1 p_3}, \quad \sin \theta_1' = \frac{\sqrt{(t_+ - t)(t - t_-)}}{2 p_1 p_3}, \quad (23e)$$

$$\cos \theta_4' = \frac{s_+ + s_- - 2s}{4 p_3 p_4}, \quad \sin \theta_4' = \frac{\sqrt{(s_+ - s)(s - s_-)}}{2 p_3 p_4}. \quad (23f)$$

Due to the delta function integration, $u$ is bound by equation (16e) and is given in terms of $s$ and $t$ by equation (A4). The additional constraints given by equations (A5–A8) also apply to equation (23a).

The result of equation (23) is identical to Baring (1987, equation 30), without the need to consider conversions of collision angles to the centre of mass frame, making this form easier to apply in general.

## 5 SINGULAR CASES OF NOTE

An observant reader may have noticed that the forms presented in equations (14) and (16) and equations (18) and (23) are not well defined at the points where their integration bounds are equivalent, i.e. $s_+ = s_-$, $t_+ = t_-$, or $u_+ = u_-$, which occur when particles are aligned with the axis of symmetry or have zero momenta. These are coordinate singularities arising from the definition of the polar axis in momentum-space spherical coordinates. Here, we will describe how to deal with such points under the formalism detailed in previous sections by giving specific examples of the emission spectrum for three cases of particular use: the cosmic ray, lab, and collider.

### 5.1 Cosmic ray case

Cosmic rays may be considered to be plain parallel rays incident on to the Earth's atmosphere. As such, their momentum vector $\boldsymbol{p}_3$ is best described as being aligned to the polar axis travelling downwards, i.e. $\theta_3 = \pi$. Given the cosmic rays are incident from a single direction, their emission spectra will in general be axisymmetric. However, if treated using equation (16), the $s, t, u$ integrals in this case are not well defined as $s_+ = s_-$ and $t_+ = t_-$, resulting in an unphysical value of zero for the emission spectrum for all output cases if numerical integration is used. To treat this point accurately, begin with the general case equation (13), replace components with their axisymmetric counterparts (as described at the beginning of Section 3) with the inclusion of a delta function as $f(p_3, \theta_3) = f(p_3)\delta(\cos \theta_3 + 1)$. With the integration over the delta function taking place first, the usual change of variables can be instated to yield the result:

$$S_{\text{spe,CR}}(p_1, \theta_1) = \frac{2 p_1^2}{\pi p_1^0} \int \frac{dp_3 dp_4 d\cos\theta_4}{p_3^0 p_4^0} f(p_3) f(p_4, \theta_4)$$

$$\times \mathcal{F}_{34}^2(s') \sigma_{34|12}(s', t') \frac{\Theta\left(p_3^0 + p_4^0 - p_1^0\right)}{\sqrt{(u_+ - u')(u' - u_-)}}, \quad (24a)$$

with

$$s' = m_3^2 + m_4^2 + 2 p_3^0 p_4^0 + 2 p_3 p_4 \cos \theta_4, \quad (24b)$$

$$t' = m_3^2 + m_1^2 - 2 p_3^0 p_1^0 - 2 p_3 p_1 \cos \theta_1, \quad (24c)$$

$$u' = m_1^2 + m_2^2 + m_3^2 + m_4^2 - s' - t', \quad (24d)$$

and $u_\pm$ given by equation (16d). Note that the factor of $1 + \delta_{34}$ has been removed as the distributions of type-3 and type-4 particles have been distinguished even though they may be of the same species. For the cosmic ray case, it is also convenient to consider the population of target particles, i.e. type 4, to be isotropic. For such a case, the emission spectrum is given by

$$S_{\text{spe,CR}}(p_1, \theta_1) = \frac{p_1^2}{2\pi p_1^0} \int \frac{dp_3 dp_4}{p_3 p_4 p_3^0 p_4^0} f(p_3) f(p_4)$$
$$\times \int_{s_-}^{s_+} ds' \mathcal{F}_{34}^2(s') \sigma_{34|12}(s', t') \frac{\Theta\left(p_3^0 + p_4^0 - p_1^0\right)}{\sqrt{(u_+ - u')(u' - u_-)}}, \quad (25)$$

with $t'$ given by equation (24c), $u'$ by equation (24d), $s_\pm$ given by equation (23b), and $u_\pm$ by equation (23d).

### 5.2 Lab case

The lab case is when one of the incident particles is at rest, e.g. $\boldsymbol{p}_4 = \boldsymbol{0}$ and $f(\boldsymbol{p}_4) = \frac{n_4 \delta(p_4)}{4\pi p_4^2}$, where $n_4$ is the number density of particles of type 4. In this case, both axisymmetric and isotropic $\boldsymbol{p}_3$ cases can be considered. For the axisymmetric case, the emission spectrum is given by

$$S_{\text{spe,L}}(p_1, \theta_1) = \frac{2p_1^2}{\pi p_1^0} \frac{n_4}{m_4} \int \frac{dp_3 d\cos\theta_3}{p_3^0} f(p_3, \theta_3) \mathcal{F}_{34}^2(s')$$
$$\times \int_{t_-}^{t^+} dt \frac{\sigma_{34|12}(s', t) \Theta\left(p_3^0 + p_4^0 - p_1^0\right)}{\sqrt{(t_+ - t)(t - t_-)}}, \quad (26a)$$

with

$$s' = m_3^2 + m_4^2 + 2p_3^0 m_4, \quad (26b)$$

and $t_\pm$ given by equation (16c). For the isotropic case,

$$S_{\text{spe,L}}(p_1) = \frac{2p_1}{p_1^0} \frac{n_4}{m_4} \int \frac{dp_3}{p_3 p_3^0} f(p_3) \mathcal{F}_{34}^2(s')$$
$$\times \sigma_{34|12}(s', t') \Theta\left(p_3^0 + p_4^0 - p_1^0\right), \quad (27a)$$

with $s'$ given by equation (26b) and

$$t' = m_1^2 + m_2^2 + m_3^2 + m_4^2 - s' - u', \quad (27b)$$

with

$$u' = m_4^2 + m_1^2 - 2p_1^0 m_4. \quad (27c)$$

The value of $t'$ in equation (26a) is subject to the additional constraint of

$$t_- \leq t' \leq t_+, \quad (27d)$$

with $t_\pm$ given by equation (23c) due to one of the delta function integrations.

### 5.3 Collider case

The collider case is an extension of the cosmic ray case (Section 5.1) where both incident particles are travelling along the axis of symmetry in opposite directions, i.e. $f(p_3, \theta_3) = f(p_3)\delta(\cos\theta_3 + 1)$ and $f(p_4, \theta_4) = f(p_4)\delta(\cos\theta_4 - 1)$. Assuming that $p_3 \neq p_4$, the emission spectrum will be axisymmetric but not isotropic in general. The form of the emission spectrum is given by

$$S_{\text{spe,C}}(p_1, \theta_1) = \frac{4p_1^2}{p_1^0} \int \frac{dp_3 dp_4}{p_3^0 p_4^0} f(p_3) f(p_4) \mathcal{F}_{34}^2(s')$$
$$\times \sigma_{34|12}(s', t') \delta\left(s' + t' + u' - \sum_i m_i^2\right) \Theta\left(p_3^0 + p_4^0 - p_1^0\right) \quad (28a)$$

where

$$s' = m_3^2 + m_4^2 + 2p_3^0 p_4^0 + 2p_3 p_4, \quad (28b)$$

$$t' = m_3^2 + m_1^2 - 2p_3^0 p_1^0 - 2p_3 p_1 \cos\theta_1, \quad (28c)$$

and

$$u' = m_4^2 + m_1^2 - 2p_4^0 p_1^0 - 2p_4 p_1 \cos\theta_1. \quad (28d)$$

The remaining delta function indicates that $p_1$ and $\theta_1$ are no longer independent variables.

If $p_3 = p_4$ and it is assumed that incident particles are travelling at a known momentum, then $f(p_3, \theta_3) = n_3\delta(p_3' - p_3)\delta(\cos\theta_3 + 1)$ and $f(p_4, \theta_4) = n_4\delta(p_4 - p_3)\delta(\cos\theta_4 - 1)$ and the emission spectrum is then given by

$$S_{\text{spe,C}}(p_1, \theta_1) = \frac{4p_1^2}{p_1^0} \frac{n_3 n_4}{p_3^0 \sqrt{p_3^2 + m_4^2}} \mathcal{F}_{34}^2(s') \sigma_{34|12}(s', t')$$
$$\times \delta\left(s' + t' + u' - \sum_i m_i^2\right) \Theta\left(p_3^0 + \sqrt{p_3^2 + m_4^2} - p_1^0\right) \quad (29a)$$

with

$$s' = m_3^2 + m_4^2 + 2p_3^0 \sqrt{p_3^2 + m_4^2} + 2p_3^2, \quad (29b)$$

$$u' = m_4^2 + m_1^2 - 2\sqrt{p_3^2 + m_4^2} p_1^0 - 2p_3 p_1 \cos\theta_1, \quad (29c)$$

and $t'$ given by equation (28c).

## 6 EXAMPLE: ELECTRON–POSITRON ANNIHILATION EMISSION SPECTRUM

The angle-averaged (isotropic) emissivity $\mathcal{E}_{\text{iso}}$ is a measure of the rate of emission of particles of particular momentum from an interaction as a function of the momentum/energy of the initial state. This was analytically generated by Svensson (1982, equations 41 and 55–58)[8] for the particular case of photon emissions from electron–positron annihilation, i.e. $e^+e^- \rightarrow \gamma\gamma$. The process of deriving the emissivity in this case involved significant algebraic manipulation and careful analysis of boundary conditions. As an example use case of the framework presented in Section 4, Svensson's work can be replicated numerically with ease and may further be extended to the axisymmetric case using the methods presented in Section 3.

The angle-averaged emissivity is related in general to equation (23) by

$$S_{\text{spe}}(p_1) = \int dp_3 dp_4 \ f(p_3) f(p_4) \mathcal{E}_{\text{iso}}, \quad (30)$$

$$\mathcal{E}_{\text{iso}} \equiv \frac{1}{1 + \delta_{34}} \frac{p_1}{4\pi p_1^0} \frac{1}{p_3^2 p_4 p_3^0 p_4^0} \int_{s_-}^{s_+} ds \ \mathcal{F}_{34}^2(s)$$
$$\times \int_{t_-}^{t_+} dt \frac{\sigma_{34|12}(s, t)}{\sqrt{(u_+ - u)(u - u_-)}} \Theta\left(p_3^0 + p_4^0 - p_1^0\right). \quad (31)$$

The Lorentz-invariant differential cross-section for the electron–positron annihilation to two photons is given by (Berestetskii, Lifshits & Pitaevskii 1982, equation 88.4)[9]

$$\sigma_{e^+e^-|\gamma\gamma}(s, t) = \frac{-3\sigma_T m_e^2}{s(s - 4m_e^2)} \left[ \left(\frac{m_e^2}{t - m_e^2} + \frac{m_e^2}{m_e^2 - s - t}\right)^2 \right.$$
$$+ \left(\frac{m_e^2}{t - m_e^2} + \frac{m_e^2}{(m_e^2 - s - t)}\right) - \frac{1}{4} \left. \left(\frac{t - m_e^2}{m_e^2 - s - t} + \frac{m_e^2 - s - t}{t - m_e^2}\right) \right], \quad (32)$$

---

[8]The angle-averaged emissivity is written as $\overline{v \frac{d\sigma}{dk}}$ in the original source.
[9]Care is to be taken with this source as it makes the simplification that $d(-t) = dt$.

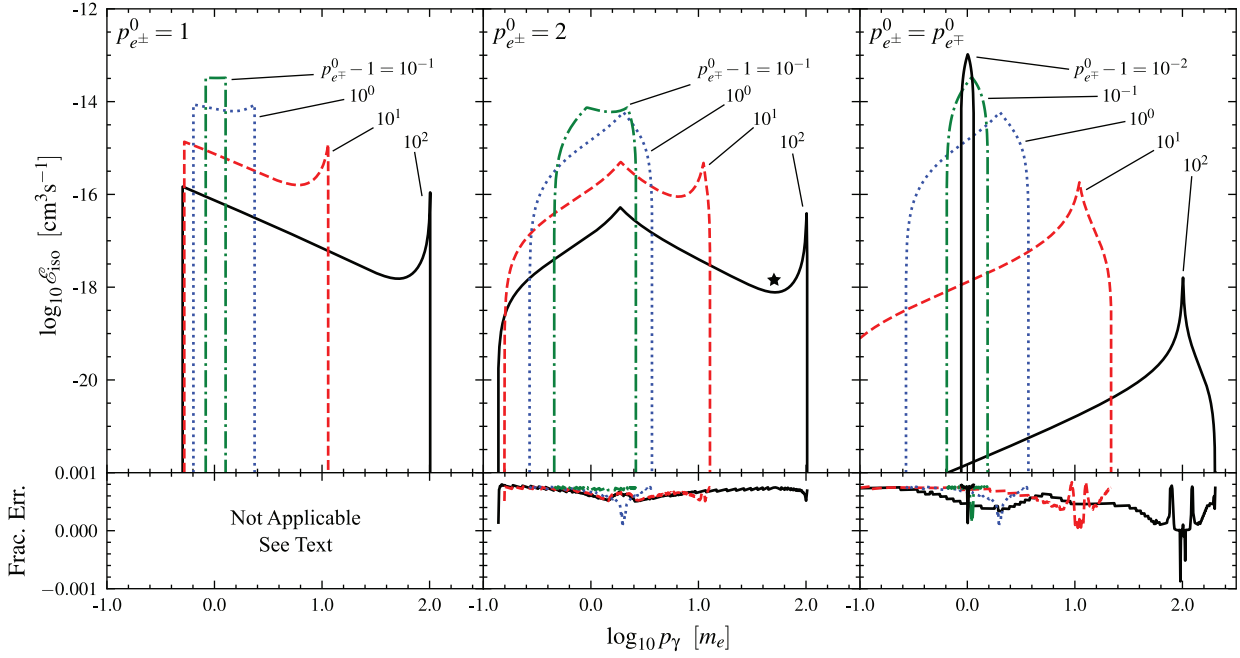

**Figure 1.** Electron–positron annihilation angle-averaged emissivity as a function of emergent photon and pair energy, $p_\gamma$ and $p_{e\pm}^0$, respectively. This figure is a reproduction of Svensson (1982, fig. 5) using the analytical form of equation (26) for the first panel and numerical integration of equation (31) for the latter two panels including the fractional error compared with Svensson (1982, equations 55–58). The angular dependence of the line with a ★ is explored in Fig. 2.

where $\sigma_T$ is the Thompson scattering cross-section and $m_e$ is the electron mass. By numerically integrating[10] equation (31) using equation (32), Svensson's work is reproduced in Fig. 1. The notation of Svensson's work has been converted to be consistent with this work. The incident electron/positron energy is denoted by $p_{e\pm}^0$ or $p_{e\mp}^0$, with the use of $\pm$ there to represent the exchange symmetry of the emissivity. The fraction errors shown in Fig. 1 are within the precision goal of the numerical scheme, demonstrating the reliability and use of equation (23) over tedious analytical integration. However, the fractional errors do tend to be positive, suggesting that numerical integration is underestimating the actual value, and can fluctuate around maximal points of the emissivity; hence, a small amount of care must be taken around these points. The emissivity in the first panel, where $p_{e\pm}^0 = 1$, is a singular case of equation (23) and has instead been calculated using equation (26). This form is analytical, and when applied to the case here of electron–positron annihilation is equivalent to Svensson (1982, equation 41).

Using the equations derived as part of this work, Svensson's isotropic emissivity may be extended to the axisymmetric case with an angle-dependent emissivity related to equation (16):

$$S_{\text{spe}}(p_1, \theta_1) = \int dp_3 dp_4 \int d\cos\theta_3 d\cos\theta_4 \qquad (33)$$
$$\times f(p_3, \theta_3) f(p_4, \theta_4) \mathcal{E}_{\text{axi}},$$

$$\mathcal{E}_{\text{axi}} \equiv \frac{1}{1+\delta_{34}} \frac{p_1^2}{2\pi^3 p_1^0 p_3^0 p_4^0} \int_{s_-}^{s_+} ds \, \frac{\mathcal{F}_{34}^2(s)}{\sqrt{(s_+ - s)(s - s_-)}}$$
$$\times \int_{t_-}^{t_+} dt \, \frac{\sigma_{34|12}(s, t)}{\sqrt{(t_+ - t)(t - t_-)}} \frac{\Theta(p_3^0 + p_4^0 - p_1^0)}{\sqrt{(u_+ - u)(u - u_-)}}. \qquad (34)$$

The spectral lines of Fig. 1 may be expanded using equation (34) to reveal spectral 'islands', which display dependence on momentum and angle from the axis of symmetry. A single spectral line from Fig. 1 is selected (labelled with a ★) and expanded in Fig. 2 for a specific set of incident angles. The incident electron–positron pair is travelling within two cones in momentum space, one with $p_{e\pm}^0 = 2$ at an angle of $\theta_\pm = 3\pi/4$ while the other at angle of $\theta_\mp = \pi/4$ with a higher energy of $p_{e\mp}^0 - 1 = 10^2$. Fig. 2 shows the angle-dependent axisymmetric emissivity of emitted photons as a function of $p_\gamma$ and $\theta_\gamma$. The emissivity peaks in regions closely located around what would be considered a 'glancing' collision, where outgoing photons match the trajectory of the incoming pair. This effect lines up with the peaks of the isotropic emissivity observed in Fig. 1. The breadth of the emissivity in emitted photon energy is narrower than the isotropic case, being truncated at low momenta, with the smallest momentum being around $10^0$, compared with $\approx 10^{-0.8}$, indicating that given the incident angles, these photon momenta are no longer physically obtainable. Two sweeping ridges can be observed in Fig. 2, a feature not present in the isotropic case, which relate to likely pairs of emitted photon states. Consider two photons emitted at the peaks of the emissivity: As one, e.g. the higher energy photon, deviates away from the peak in phase space, the other must travel in order to conserve momentum and energy. If it travels down the peak to the north-west of Fig. 2, its cone of emission (remembering that these photons are emitted symmetrically about the axis) has converged closer to the axis of symmetry and it has lost energy. To conserve momentum, the lower energy photon must begin to travel north-east in order to broaden its cone of emission and gain energy.

---

[10]Numerical integration is performed using a GlobalAdaptive scheme in Mathematica version 13.2, with a PrecisionGoal of 3.

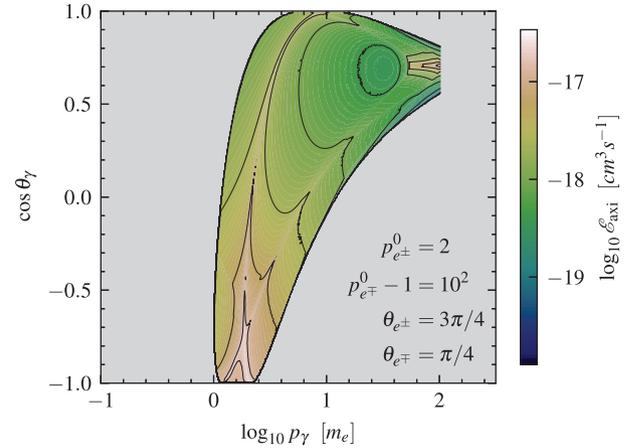

**Figure 2.** Electron–positron annihilation angle-dependent emissivity for $p_{e\pm}^0 = 2$, $p_{e\mp}^0 = 1 + 10^2$, $\theta_{e\pm} = \frac{3\pi}{4}$, and $\theta_{e\mp} = \frac{\pi}{4}$.

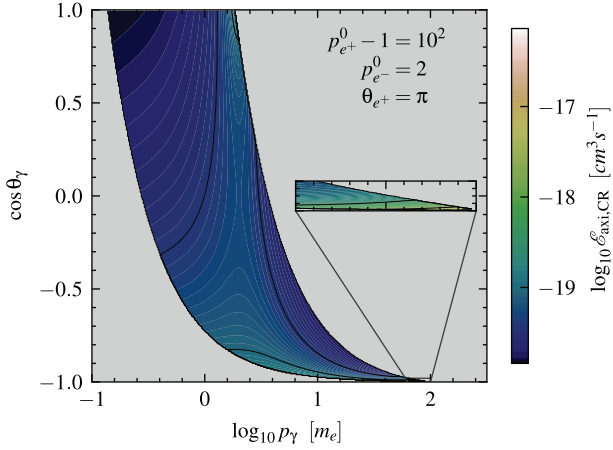

**Figure 3.** Angle-dependent emissivity of photons from high-energy cosmic ray positrons at an angle of $\theta_{e^+} = \pi$ at an energy of $p^0_{e^+} = 1 + 10^2$ incident on an isotopically distributed, low-energy population of electrons with $p^0_{e^-} = 2$.

For an alternate example of photon emissivity, Fig. 3 presents the particular case of high-energy positrons ($p^0_{e^+} - 1 = 10^2$) incident along the axis of symmetry, on an isotropic population of low-energy electrons with energy $p^0_{e^-} = 2$. This emissivity is generated from equation (25), as in a cosmic ray case, and peaks around the location of incident positron; however, unlike the case in Fig. 2, the electrons have no preferred orientation in momentum space, leading to a broadened out secondary peak of photon emission around the electron energy ($p_\gamma \approx 2$). It may also be noted that even though the incident electrons are isotropic, the emergent photons are not, even at the electron's energy where emission angles of greater than $\theta_\gamma \approx 150°$ are unphysical. This may have utility in calculating the angular distribution of particles in air shower simulations.

## 7 CONCLUSION

New computational forms for emission and absorption spectra have been generated for relativistic, binary interactions with varying levels of anisotropy. These take the form of multidimensional integrals, of minimal dimensionality. Their use is analysed by comparison with previously discovered integral forms and is demonstrated to agree to numerical accuracy to the specific case of isotropic electron–positron annihilation, for which an analytically integrated form of the emission spectrum is known.

The specific case of electron–positron annihilation is briefly expanded upon to demonstrate the potential use of the new computational forms for axisymmetric emission spectrum presented in this work. Two cases are examined: photon emissions from an axisymmetric state of electron–positron pairs, and a cosmic ray case where high-energy positrons are incident on an isotropic distribution of low-energy electrons. A spectral island is generated for each, denoting the emissivity of photons as a function of momenta and angle from the axis of symmetry.

The computational forms presented here are expected to be useful in work requiring accurate description of collisional systems with anisotropic processes. Further papers will explore such scenarios, with an application to dynamical modelling of particle production and high-energy emissions in astrophysical jets.

## DATA AVAILABILITY

The scripts for generating data underlying this article will be shared on reasonable request to the corresponding author.

## ACKNOWLEDGEMENTS

CNE acknowledges an STFC studentship ST/X508664/1. GC acknowledges support from STFC grants ST/V006355/1, ST/V001477/1, and ST/S002952/1 and from Exeter College, Oxford.

## REFERENCES

Aksenov A. G., Milgrom M., Usov V. V., 2004, ApJ, 609, 363
Aksenov A. G., Ruffini R., Vereshchagin G. V., 2007, Phys. Rev. Lett., 99, 125003
Aksenov A. G., Ruffini R., Vereshchagin G. V., 2010, Phys. Rev. E, 81, 046401
Baring M., 1987, MNRAS, 228, 681
Berestetskii V. B., Lifshits E. M., Pitaevskii L. P., 1982, Quantum Electrodynamics, 2nd edn. Butterworth-Heinemann, Oxford, UK
Cercignani C., 2002, The Relativistic Boltzmann Equation: Theory and Applications, Progress in Mathematical Physics, Vol. 22. Birkhäuser, Basel
Cerruti M. et al., 2021, Proc. Sci., 37th Int. Cosm. Ray Conf. (ICRC2021): MM - Multi-Messenger. SISSA, Trieste, PoS#979
Coppi P. S., Blandford R. D., 1990, MNRAS, 245, 453
Dermer C. D., 1984, ApJ, 280, 328
Dermer C. D., 1985, ApJ, 295, 28
Dermer C. D., 1986, ApJ, 307, 47
Groot S. R., Leeuwen W. A., Weert C. G., 1980, Relativistic Kinetic Theory: Principles and Applications. North-Holland Publ. Co., New York
Källén G., 1964, Elementary Particle Physics, A-W Series in Advanced Physics. Addison-Wesley Publ. Co., Boston
Klinger M. et al., 2023, preprint (arXiv:2312.13371)
Lucchini M. et al., 2022, MNRAS, 517, 5853
Potter W. J., 2018, MNRAS, 473, 4107
Stepney S., Guilbert P. W., 1983, MNRAS, 204, 1269
Svensson R., 1982, ApJ, 258, 321
Weaver T. A., 1976, Phys. Rev. A, 13, 1563
Zacharias M., Reimer A., Boisson C., Zech A., 2022, MNRAS, 512, 3948

## APPENDIX A: MANDELSTAM VARIABLES AND THEIR BOUNDS

The Mandelstam variables $s, t, u$ for the binary collision $12 \rightleftharpoons 34$ are defined by

$$\begin{aligned} s &= (p^\mu_1 + p^\mu_2)^2 = m^2_1 + m^2_2 + 2p^\mu_1 p_{2\mu}, \\ &= (p_{3\mu} + p^\mu_4)^2 = m^2_3 + m^2_4 + 2p^\mu_3 p_{4\mu}, \end{aligned} \quad (A1)$$

$$t = (p_1^\mu - p_3^\mu)^2 = m_1^2 + m_3^2 - 2p_1^\mu p_{3\mu},$$
$$= (-p_{2\mu} + p_4^\mu)^2 = m_2^2 + m_4^2 - 2p_2^\mu p_{4\mu}, \quad (A2)$$

$$u = (p_1^\mu - p_4^\mu)^2 = m_1^2 + m_4^2 - 2p_1^\mu p_{4\mu},$$
$$= (-p_{2\mu} + p_3^\mu)^2 = m_2^2 + m_3^2 - 2p_2^\mu p_{3\mu}. \quad (A3)$$

They are related by the identity

$$s + t + u = \sum_i m_i^2. \quad (A4)$$

Not all sets of Mandelstam variables represent real collisions. The physical region is defined by the constraints (Berestetskii et al. 1982):

$$(m_1 + m_2)^2 \leq s \geq (m_3 + m_4)^2, \quad (A5)$$

$$(m_1 - m_3)^2 \geq t \leq (m_4 - m_2)^2, \quad (A6)$$

$$(m_1 - m_4)^2 \geq u \leq (m_3 - m_2)^2, \quad (A7)$$

and

$$stu \geq as + bt + cu, \quad (A8)$$

where

$$ah = (m_1^2 m_2^2 - m_3^2 m_4^2)(m_1^2 + m_2^2 - m_3^2 - m_4^2), \quad (A9)$$

$$bh = (m_1^2 m_3^2 - m_2^2 m_4^2)(m_1^2 + m_3^2 - m_2^2 - m_4^2), \quad (A10)$$

$$ch = (m_1^2 m_4^2 - m_2^2 m_3^2)(m_1^2 + m_4^2 - m_2^2 - m_3^2), \quad (A11)$$

$$h = \sum_i m_i^2. \quad (A12)$$

## APPENDIX B: INTEGRATED ANISOTROPIC EMISSION SPECTRUM

The delta function in equation (13) may be eliminated by integrating $S_{\rm spe}(\boldsymbol{p}_1)$ over $p_1^2 {\rm d} p_1$. Integration is performed using the identity

$$\delta(g(x)) = \sum_i \frac{\delta(x - x_i)}{|\partial g(x)/\partial p|_{x=x_i}}, \quad (B1)$$

where $g(x)$ is some arbitrary smooth function and $x_i$ are its roots, yielding the result

$$S_{\rm spe}(\theta_1, \phi_1) = \frac{1}{1+\delta_{34}} \frac{2}{\pi} \int \frac{{\rm d}^3 \boldsymbol{p}_3}{p_3^0} \frac{{\rm d}^3 \boldsymbol{p}_4}{p_4^0} f(\boldsymbol{p}_3) f(\boldsymbol{p}_4) \mathcal{F}_{34}(s)^2$$
$$\times \sum_\pm \frac{p_\pm^2}{p_\pm^0} \sigma_{34|12}(s, t_\pm) \frac{\Theta(p_3^0 + p_4^0 - p_\pm^0)}{|\partial g/\partial p_1|_{p_1 = p_\pm}}. \quad (B2)$$

The function $g$ in this case is given by

$$g = s + t + u - m_1^2 - m_2^2 - m_3^2 - m_4^2, \quad (B3)$$

with partial derivative

$$\frac{\partial g}{\partial p_1} = \frac{-2}{p_1^0} \left( p_3^0 p_1 - p_1^0 p_3 \cos \Theta_{13} + p_4^0 p_1 - p_1^0 p_4 \cos \Theta_{14} \right), \quad (B4)$$

where $\cos \Theta_{1i} = \hat{\boldsymbol{p}}_1 \cdot \hat{\boldsymbol{p}}_i$ and $p_\pm$ are the roots of $g$. In general, there are two roots of $g$ for each set of input variables; however, care must be taken as one of the roots may be negative, physically corresponding to positive momentum magnitude with mirrored direction, i.e. $\theta_1 \to \pi - \theta_1$ and $\phi_1 \to \phi_1 + \pi$.

By performing this integration, the magnitude of $\boldsymbol{p}_1$ is no longer an independent variable, instead is fully determined by the roots of $g$ which depends on the values of $\boldsymbol{p}_{3,4}, \theta_1$, and $\phi_1$.

This paper has been typeset from a T$_{\rm E}$X/L$^{\rm A}$T$_{\rm E}$X file prepared by the author.